\documentclass[fleqn,usenatbib,onecolumn]{mnras}

\usepackage{newtxtext,newtxmath}

\usepackage[T1]{fontenc}
\usepackage{ae,aecompl}

\usepackage[british]{babel}

\usepackage{graphicx}	
\usepackage{amsmath}	

\usepackage{siunitx}    

\DeclareSIUnit\solarMass{\mbox{$M_\odot$}}

\title{Modelling neutron star mountains in relativity}

\author[F. Gittins and N. Andersson]{
Fabian Gittins\thanks{E-mail: f.w.r.gittins@soton.ac.uk}
and Nils Andersson
\\
Mathematical Sciences and STAG Research Centre, 
University of Southampton, Southampton SO17 1BJ, United Kingdom
}

\date{}

\pubyear{2021}

\begin{document}
\label{firstpage}
\pagerange{\pageref{firstpage}--\pageref{lastpage}}
\maketitle

\begin{abstract}
Rapidly spinning, deformed neutron stars have long been considered potential 
gravitational-wave emitters. However, so far only upper limits on the size of 
the involved quadrupole deformations have been obtained. For this reason, it is 
pertinent to ask how large a mountain can be before the neutron star crust 
fractures. This is the question we consider in this paper, which describes how 
mountains can be calculated in relativistic gravity. Formally, this is a 
perturbative calculation that requires a fiducial force to source the mountain. 
Therefore, we consider three simple examples and increase their deforming 
amplitudes until the crust yields. We demonstrate how the derived mountains 
depend on the equation of state by considering a range of models obtained from 
chiral effective field theory. We find that the largest mountains depend 
sensitively on both the mechanism that sources them and the nuclear-matter 
equation of state.
\end{abstract}



\section{Introduction}

With the recent detections of gravitational waves from binary neutron star 
mergers \citep{2017PhRvL.119p1101A, 2020ApJ...892L...3A} and sensitivity 
improvements of the instruments, there is much anticipation for observing 
signals from rapidly rotating neutron stars for the first time. Provided a 
spinning neutron star is deformed in a non-axisymmetric way, it will 
continuously emit gravitational radiation. These quadrupole deformations are 
known as \textit{mountains}. Historically, these systems have been of interest 
since the presence of mountains -- resulting in gravitational-wave torques that 
would spin down the stars -- could play an important role in determining the 
spin equilibrium for low-mass X-ray binaries 
\citep*{1998ApJ...501L..89B, 1999ApJ...516..307A}. Thus, providing an 
explanation for their observed spin distribution, in particular, the lack of 
sub-millisecond pulsars \citep*{2017ApJ...850..106P, 2019MNRAS.488...99G}.

Although, we are yet to be graced with a confirmed gravitational-wave detection 
of a spinning neutron star, there has been a significant effort to look for 
signals from individual pulsars 
\citep{2004PhRvD..69h2004A, 2005PhRvL..94r1103A, 2007PhRvD..76d2001A, 
2008ApJ...683L..45A, 2010ApJ...713..671A, 2017PhRvD..95l2003A, 
2017PhRvD..96l2006A, 2017ApJ...839...12A, 2017ApJ...847...47A, 
2018PhRvL.120c1104A, 2019PhRvD..99l2002A, 2019PhRvD.100l2002A, 
2019ApJ...879...10A, 2020arXiv201212926T, 2020ApJ...902L..21A, 
2011PhRvD..83d2001A, 2011ApJ...737...93A, 2014ApJ...785..119A, 
2015PhRvD..91b2004A, 2015PhRvD..91f2008A, 2021ApJ...906L..14Z}, 
accompanied by wide-parameter searches for unknown systems 
\citep{2005PhRvD..72j2004A, 2007PhRvD..76h2001A, 2008PhRvD..77b2001A, 
2009PhRvD..79b2001A, 2016PhRvD..94d2002A, 2017PhRvD..96f2002A, 
2018PhRvD..97j2003A, 2019PhRvD.100b4004A, 2012PhRvD..85b2001A, 
2013PhRvD..87d2001A, 2020PhRvL.125q1101D, 2021PhRvD.103f3019D, 
2021ApJ...909...79S}. 
These surveys, through non-detections, have so far only yielded upper limits on 
the size of possible deformations. Looking at more recent surveys: 
\citet{2020ApJ...902L..21A} were able to constrain the ellipticities for the 
millisecond pulsars PSR J0437-4715 and PSR J0711-6830 to be less than 
\num{e-8}, while \citet{2020arXiv201212926T} obtained a constraint of less than 
\num{3e-5} for the young pulsar PSR J0537-6910. Indeed, the general trend is 
that the upper limits of ellipticities of neutron stars are being constrained to 
lower and lower values. One might think that a pattern is emerging. Furthermore, 
it has been argued by \citet{2018ApJ...863L..40W} that there exists a 
\textit{minimum} ellipticity for millisecond pulsars of 
$\epsilon \approx \num{e-9}$. In the same vein, \citet{2019MNRAS.488...99G} have 
shown that the features of the spin distribution of accreting millisecond 
pulsars can be explained by stars with quadrupole deformations of the order of 
$\epsilon \approx \num{e-9}$. It is worth noting that there is only one order of 
magnitude difference between the smallest established upper limit and \num{e-9}, 
implying the observational constraints are approaching this minimum ellipticity.

The amplitude of the radiation scales with the size of the mountain. Thus, it 
is a natural question to ask how large a quadrupole deformation can the neutron 
star crust support? Indeed, this question has been the subject of a number of 
studies over the past two decades 
\citep*{2000MNRAS.319..902U, 2006MNRAS.373.1423H, 2013PhRvD..88d4004J, 
2021MNRAS.500.5570G}. For some time, the standard method to solving this 
problem has been to follow \citet{2000MNRAS.319..902U}, the first to carry out 
such a maximum-mountain calculation in Newtonian gravity with the aid 
of the Cowling approximation. By ensuring the crust was maximally strained at 
every point they obtained the strain field of the crust and could, thus, 
calculate the mountain it supported. Later, this calculation was generalised to 
relativistic gravity by \citet{2013PhRvD..88d4004J}, while also relaxing the 
Cowling approximation. However, it has been recently pointed out that there are 
inconsistencies with all previous studies \citep{2021MNRAS.500.5570G}. In 
particular, there are issues with the boundary conditions of the problem if one 
is to demand that the crust is maximally strained \textit{at every point}. To 
this end, \citet{2021MNRAS.500.5570G} introduced a novel scheme for calculating 
neutron star mountains. This scheme requires a fiducial force that sources the 
mountain, which is contrary to the \citet{2000MNRAS.319..902U} approach, where 
the force is, in some sense, hidden in the calculation. An important consequence 
of this work is that the maximum size of a mountain strongly depends on its 
formation. Therefore, evolutionary calculations will be required in order to 
make progress on the problem.

The aim of this work is to translate the scheme of \citet{2021MNRAS.500.5570G} 
from Newtonian gravity to relativity -- an important step towards realism and 
greater accuracy which enables us to explore more realistic equations of state. 
In doing so, we are able to consider how the size of neutron star mountains 
depend on the description of nuclear matter.

We dedicate Section~\ref{sec:Calculating} to a discussion on how one obtains 
the multipole moments of a body in relativistic gravity and, then, we briefly 
review the mountain scheme of \citet{2021MNRAS.500.5570G}. In 
Section~\ref{sec:Models}, we provide the formalism that describes non-spherical 
stellar models with and without crusts. This is carried out in a perturbative 
manner, where the non-spherical, even-parity perturbations are constructed on 
top of a spherical, fluid background. It is evident from existing results that 
the mountain problem is well within the linear perturbation regime. We also 
carefully detail the boundary conditions of the problem. We consider a selection 
of forces to produce mountains in Section~\ref{sec:Source} for a sequence of 
stars with the same realistic equation of state. We also study how the size of 
the mountains vary with a collection of equations of state from chiral effective 
field theory in Section~\ref{sec:Dependence}. To close, we conclude and discuss 
where future progress will be found for mountain calculations in 
Section~\ref{sec:Conclusions}.

We use the metric signature $(-, +, +, +)$ and work in geometric units, where 
$G = c = 1$. We adopt the Einstein summation convention where repeated indices 
denote a summation. We use early Latin characters $a, b, c, ...$ for spacetime 
indices and later characters $i, j, k, ...$ for spatial indices. We reserve the 
characters $l$ and $m$ for spherical-harmonic modes. We use primes to denote 
derivatives with respect to the radial coordinate. Much of the analysis will use 
perturbation theory, where we denote the Eulerian perturbation of a quantity 
using $\delta$ and the Lagrangian perturbation using $\Delta$. The two are 
related by $\Delta = \delta + \mathcal{L}_\xi$, where $\mathcal{L}_\xi$ is the 
Lie derivative along the Lagrangian displacement vector, $\xi^a$ 
\citep{1978ApJ...221..937F}.

\section{Calculating mountains}
\label{sec:Calculating}

\subsection{Multipole moments in relativity}

There is freedom in how one chooses to define the multipole moments. However, we 
must be consistent in order to compare previous Newtonian calculations 
\citep{2000MNRAS.319..902U, 2006MNRAS.373.1423H, 2021MNRAS.500.5570G} with this 
calculation in relativity. The quadrupole we obtain matches the definition used 
in the relativistic calculation of \citet{2013PhRvD..88d4004J}.

We treat multipole moments as departures from perfect spherical symmetry. In 
Newtonian gravity, consider a star that is deformed away from sphericity 
according to the mass-density perturbation, $\delta \rho$. The multipole moments 
of the star are then defined as 
\begin{equation}
    Q_{l m} \equiv \int_V \delta \rho r^l Y_{l m}^* d V, 
\label{eq:MultipoleNewtonian2}
\end{equation}
where $(l, m)$ is the order of the spherical-harmonic mode, 
$Y_{l m}(\theta, \phi)$, and the star denotes a complex conjugate. One is free 
to decompose the angular behaviour of an arbitrary function using spherical 
harmonics, e.g., the perturbation of the mass density may be written as 
$\delta \rho(r, \theta, \phi) =  \sum_{l = 0}^\infty \sum_{m = -l}^l 
\delta \rho_{l m}(r) Y_{l m}(\theta, \phi)$. Therefore, 
(\ref{eq:MultipoleNewtonian2}) can be expressed as 
\begin{equation}
    Q_{l m} = \int_0^R \delta \rho_{l m}(r) r^{l + 2} dr,
\label{eq:MultipoleNewtonian}
\end{equation}
where $R$ is the radius of the background spherical star. By combining the 
perturbed Poisson's equation,
\begin{equation}
    \nabla^2 \delta \Phi = 4 \uppi \delta \rho,
\end{equation}
where $\delta \Phi$ is the perturbation of the gravitational potential with 
boundary conditions (assuming that the density vanishes at the star's surface)
\begin{equation}
    \delta \Phi_{l m}(0) = 0, \qquad 
        R \delta \Phi_{l m}'(R) = - (l + 1) \delta \Phi_{l m}(R),
\label{eq:BoundaryConditions}
\end{equation}
with the multipole definition (\ref{eq:MultipoleNewtonian}), one can show, for 
$r \geq R$,
\begin{equation}
    \delta \Phi_{l m}(r) = - \frac{4 \uppi}{2 l + 1} \frac{Q_{l m}}{r^{l + 1}}.
\end{equation}
Therefore, provided one is able to find the perturbed potential in the exterior 
of the star, one can straightforwardly read off the multipole moments, 
$Q_{l m}$.

In general relativity, the spacetime metric, $g_{a b}$, replaces the Newtonian 
potential, $\Phi$. In the far-field exterior of the star, $r \gg R$, one obtains 
the Newtonian limit, 
\begin{equation}
    - \frac{1 + g_{t t}}{2} = \Phi + \chi,
\label{eq:Limit}
\end{equation}
where $\chi$ is the corresponding potential to the force that sources the 
deformation. It should be noted that the deforming potential, $\chi$, is 
necessary in order to induce a non-spherical deformation 
\citep{2021MNRAS.500.5570G}. Equation~(\ref{eq:Limit}) illustrates an important 
point. Although in Newtonian gravity one can analyse the gravitational and 
deforming potentials separately, such a neat decoupling does not necessarily 
exist in relativity. In relativity, one has the metric that accounts for all the 
potentials acting on the body. For this reason, in some cases, it is not 
possible to disentangle the separate potentials from the metric and ambiguities 
arise.%
\footnote{To be specific, ambiguities occur when different potentials mix in 
powers of $r$. Much of the work concerning separating multiple gravitational 
effects in relativity has come from studies on tidal deformations 
\citep[see, e.g.,][]{2018CQGra..35h5002G}.}
However, no ambiguities occur when one considers pure multipoles of order $l$. 
Indeed, we will focus on pure multipoles of order $(l, m)$. In this case, for 
$r \geq R$, we have [cf. a variation of (\ref{eq:Limit})]
\begin{equation}
    - \frac{h_{t t}}{2} = 
        - \frac{4 \uppi}{2 l + 1} B_l(r) \frac{Q_{l m} Y_{l m}}{r^{l + 1}} 
        + \chi_{l m}(r) Y_{l m},
\label{eq:MultipoleRelativity}
\end{equation}
where $h_{a b} \equiv \delta g_{a b}$ is the linearised metric, $B_l(r)$ is a 
function that goes to unity in the Newtonian limit 
\citep[this is given as $B_1$ in Table I of][]{2009PhRvD..80h4018B} and 
we treat the deforming potential, $\chi$, as a first-order parameter. 
Equation~(\ref{eq:MultipoleRelativity}) is obtained by looking for a solution 
to the perturbed Einstein equations in the exterior of the star.%
\footnote{We note that (\ref{eq:MultipoleRelativity}) differs slightly, by a 
constant factor, from equivalent expressions in the tidal-deformation literature 
[see, e.g., equation~(1) in \citet{2008ApJ...677.1216H}]. This is due to the 
different convention for defining the multipole moments originating from 
\citet{1980RvMP...52..299T} [cf. their equation~(5.27a) with 
(\ref{eq:MultipoleNewtonian2})].}
(Note that $\chi$ need not be a solution to the vacuum Einstein equations. It 
simply corresponds to a force that deforms the shape of the star.) Since we are 
interested in calculating the quadrupole moment, $Q_{2 2}$, it will be 
sufficient for this work to focus on the $(l, m) = (2, 2)$ mode, so we will 
henceforth neglect the mode subscript on the perturbation variables.

\subsection{Mountain scheme}
\label{sec:Scheme}

We briefly summarise the method for calculating mountains introduced by 
\citet{2021MNRAS.500.5570G}. Contrary to the \citet{2000MNRAS.319..902U} 
approach that begins with the strain, the scheme of \citet{2021MNRAS.500.5570G} 
instead requires a description of the deforming force.

Consider two stars: (i) a purely fluid star (star A) and (ii) a star with a 
crust (star B). Both stars begin with the same spherical shape and the crust of 
star B is relaxed in this shape. The two stars are subjected to the same 
$(l, m) = (2, 2)$ force that deforms them into non-spherical shapes. As star B 
is deformed, strain is built up in the crust as the stresses resist the change 
in shape. The two stars, supported by the same force, are then subtracted from 
one another to cancel out the force.%
\footnote{Note that the two stars that are subtracted from one another are 
called star A and star C in \citet{2021MNRAS.500.5570G}.}
This procedure provides the perturbation quantities that connect a spherical 
background to a star with a mountain supported by elastic stresses. Provided a 
description of the deforming force, one can then calculate all the relevant 
information about the star with a mountain. It was shown in 
\citet{2021MNRAS.500.5570G} how this scheme does, in fact, produce the same 
non-spherical star with a mountain as considered in previous mountain 
calculations. The advantage to this new scheme is that one has full control over 
the boundary conditions of the problem. The disadvantage is the necessity to 
prescribe the force, which should ideally take into account some (presently) 
poorly understood aspects of neutron star physics.

With this scheme for computing mountains, there is a natural way to obtain the 
largest quadrupole for a given force: the amplitude of the force is increased 
until a point in the crust of star B reaches breaking strain. The obtained 
mountains can then be readily compared with previous estimates. 

\citet{2021MNRAS.500.5570G} considered several examples for the form of the 
deforming force and found that their mountains were between a factor of a few to 
two orders of magnitude smaller than the results from previous maximum-mountain 
calculations. The reason for this difference lies in the fact that the crust 
breaks at a point in the scheme of \citet{2021MNRAS.500.5570G}, rather than 
throughout the crust.

\section{Stellar models}
\label{sec:Models}

In order to implement the scheme described in Section~\ref{sec:Scheme}, we need 
to construct pairs of relativistic stellar models with and without crusts. We 
will focus on non-rotating stars and assume that stars A and B differ to a 
spherically symmetric star in a perturbative way. This enables us to generate 
the stars through perturbations on a spherical background. We use the static 
perturbation formalism provided in \citet*{2020PhRvD.101j3025G} to build the 
stars.

The fact that star B is relaxed in a spherical shape provides significant 
simplification: the crust manifests itself only at the linear perturbation 
level. In order to model the crust, we will assume that it is well approximated 
as an elastic solid. Star B will be partitioned into three layers: a fluid core, 
an elastic crust and a fluid ocean.

\subsection{Background}

In general relativity, a stellar configuration -- characterised by a spacetime 
metric, $g_{a b}$, with energy density $\varepsilon$ and isotropic pressure $p$ 
as measured by an observer with four-velocity $u^a$ -- is a solution 
$(\varepsilon, p, u^a, g_{a b})$ to the Einstein field equations,
\begin{equation}
    G_a^{\hphantom{a} b} = 8 \uppi T_a^{\hphantom{a} b},
\label{eq:Einstein}
\end{equation}
where $G_a^{\hphantom{a} b}$ is the Einstein tensor and $T_a^{\hphantom{a} b}$ 
is the stress-energy tensor, supplemented by an equation of state,
\begin{equation}
    p = p(\varepsilon),
\label{eq:State}
\end{equation}
which we have assumed to be barotropic. The equilibrium of a non-rotating, 
spherically symmetric, fluid star is described by the line element of the form,
\begin{equation}
    ds^2 = g_{a b} dx^a dx^b = - e^\nu dt^2 + e^\lambda dr^2 + r^2 (d\theta^2 
        + \sin^2 \theta \, d\phi^2),
\end{equation}
where $\nu$ and $\lambda$ are metric functions, with the stress-energy tensor,
\begin{equation}
    T_a^{\hphantom{a} b} = (\varepsilon + p) u_a u^b 
        + p \delta_a^{\hphantom{a} b} 
        = \varepsilon u_a u^b + p \perp_a^{\hphantom{a} b},
    \label{eq:StressEnergy}
\end{equation}
appropriate for perfect fluids, where 
$\perp_a^{\hphantom{a} b} \equiv u_a u^b + \delta_a^{\hphantom{a} b}$ is the 
projection operator orthogonal to the four-velocity, $u^a$, which we will use 
later. Since the star is spherically symmetric, all quantities are solely 
functions of the radial coordinate, $r$. It is convenient to choose an observer 
at rest with the fluid. As the fluid is static, the four-velocity (which must be 
normalised, $u^a u_a = -1$) may be simply given by
\begin{equation}
    u^t = e^{-\nu/2}, \qquad u^i = 0.
\end{equation}

Putting this information into the field equations~(\ref{eq:Einstein}) provides 
the relativistic equations of stellar structure for a non-rotating, fluid star:
\begin{subequations}
\begin{gather}
    m' = 4 \uppi r^2 \varepsilon, \\
    p' = - \frac{1}{2} (\varepsilon + p) \nu'
\end{gather}
and 
\begin{equation}
    \nu' = \frac{2 (m + 4 \uppi r^3 p)}{r (r - 2 m)},
\end{equation}
\end{subequations}
where $m(r) = r [1 - e^{-\lambda(r)}] / 2$ is the gravitational mass enclosed 
in $r$. Along with the equation of state (\ref{eq:State}), these structure 
equations describe the spherical background.

\subsection{Fluid perturbations}

To generate star A and the fluid regions of star B, we must consider fluid 
perturbations. The perturbations are a solution 
$(\delta \varepsilon, \delta p, \delta u^a, h_{a b})$ to the linearised Einstein 
equations,
\begin{equation}
    \delta G_a^{\hphantom{a} b} = 8 \uppi \delta T_a^{\hphantom{a} b}.
\label{eq:PerturbedEinstein}
\end{equation}
As the matter is barotropic, we have 
\begin{equation}
    \delta p = c_\text{s}^2 \delta \varepsilon,
\end{equation}
where $c_\text{s}^2 \equiv dp / d\varepsilon$ is the squared sound speed. Since 
we intend to build stars with quadrupolar deformations, we work in the 
Regge-Wheeler gauge \citep{1957PhRv..108.1063R} specialised to static, 
even-parity perturbations. This gives the linearised metric,
\begin{equation}
    h_{a b} = 
    \begin{pmatrix}
        e^\nu H_0   & H_1           & 0     & 0 \\
        H_1         & e^\lambda H_2 & 0     & 0 \\
        0           & 0             & r^2 K & 0 \\
        0           & 0             & 0     & r^2 \sin^2 \theta \, K
    \end{pmatrix}
    Y_{l m},
\end{equation}
where $H_0(r)$, $H_1(r)$, $H_2(r)$ and $K(r)$ are the perturbed metric 
functions. The perturbations of the metric are coupled to the perturbations of 
the stress-energy tensor, which for fluid perturbations are given by [cf. a 
perturbation of (\ref{eq:StressEnergy})]
\begin{equation}
    \delta T_a^{\hphantom{a} b} = (\delta \varepsilon + \delta p) u_a u^b 
        + \delta p \delta_a^{\hphantom{a} b} 
        + (\varepsilon + p) (\delta u_a u^b + u_a \delta u^b).
\label{eq:PerturbedStressEnergyFluid}
\end{equation}
The perturbed quantities are to be expanded using spherical harmonics. The 
linearised four-velocity for static, even-parity perturbations is
\begin{equation}
    \delta u^t = \frac{1}{2} e^{-\nu/2} H_0 Y_{l m}, \qquad \delta u^i = 0.
\end{equation}

The perturbed Einstein equations (\ref{eq:PerturbedEinstein}) for a fluid yield 
the following system of equations:
\begin{subequations}\label{eqs:FluidPerturbations}
\begin{gather}
\label{eq:H_0Fluid}
    H_0'' + \left( \frac{2}{r} + \frac{\nu' - \lambda'}{2} \right) H_0' 
        + \left\{ \frac{2}{r^2} - [2 + l (l + 1)] \frac{e^\lambda}{r^2} 
        + \frac{4 \nu' + 2 \lambda'}{r} - \nu'^2 \right\} H_0 
        = - 8 \uppi e^\lambda (\delta \varepsilon + \delta p), \\
    \delta p = \frac{e^{-\lambda} (\nu' + \lambda')}{16 \uppi r} H_0, \\
    H_1 = 0, \\
    H_2 = H_0
\end{gather}
and 
\begin{equation}
    [l (l + 1) - 2] e^\lambda K = r^2 \nu' H_0' + [l (l + 1) e^\lambda - 2
        - r (\nu' + \lambda') + r^2 \nu'^2] H_0.
\end{equation}
\end{subequations}
Equations~(\ref{eqs:FluidPerturbations}) provide the necessary information to 
compute static, even-parity perturbations in the fluid regions of a star. At 
this point, we should note that the perturbation equations in this form do not 
necessarily include the force that sources the deformation -- unless the force 
satisfies the relativistic analogue to Laplace's equation. In this situation, 
which is effectively the tidal problem 
\citep[see, e.g.,][]{2008ApJ...677.1216H}, the perturbed metric function, $H_0$, 
contains both the gravitational and tidal potentials. When we consider examples, 
we will discuss how one may include the force.

\subsection{Elastic perturbations}

In stellar perturbation theory, it is traditional to work with the Lagrangian 
displacement vector, $\xi^a$, that connects fluid elements in the perturbed 
configuration to their positions in the background. Due to the static nature of 
the problem, one is unable to calculate the displacement for the fluid 
perturbations. However, this is different in elastic material. We will define 
the displacement vector appropriate for static, even-parity perturbations,
\begin{equation}
	\xi^a = 
	\begin{bmatrix}
		0 \\
		r^{-1} W \\
		r^{-2} V \partial_\theta \\
		(r \sin \theta)^{-2} V \partial_\phi \\
	\end{bmatrix}
	Y_{l m}, 
\end{equation}
where the functions $W(r)$ and $V(r)$ capture the radial and tangential 
displacements, respectively.

In order to compute the crust of star B, we must consider perturbations of an 
elastic solid. For an elastic material with shear modulus $\mu$, the 
perturbed anisotropic stress tensor is \citep{2019CQGra..36j5004A}
\begin{equation}
    \Delta \pi_{a b} = - 2 \mu \Delta s_{a b},
\end{equation}
where the perturbed strain tensor, $\Delta s_{a b}$, is given by
\begin{equation}
    2 \Delta s_{a b} = \left( \perp^c_{\hphantom{c} a} \perp^d_{\hphantom{d} a} 
        - \frac{1}{3} \perp_{a b} \perp^{c d} \right) \Delta g_{c d}
\end{equation}
and the perturbation to the metric is 
$\Delta g_{a b} = h_{a b} + \nabla_a \xi_b + \nabla_b \xi_a$. For an unstrained 
background, we find
\begin{equation}
    \delta \pi_a^{\hphantom{a} b} = - \mu 
        \left( \perp^c_{\hphantom{c} a} \perp^{d b} 
        - \frac{1}{3} \perp_a^{\hphantom{a} b} \perp^{c d} \right) 
        \Delta g_{c d}.
\label{eq:PerturbedStressEnergyElastic}
\end{equation}
In the elastic crust, the linearised stress-energy 
tensor includes the fluid contribution (\ref{eq:PerturbedStressEnergyFluid}) 
summed with the elastic stress tensor (\ref{eq:PerturbedStressEnergyElastic}). 
For the application of the boundary conditions, it is convenient to define the 
following variables that are related to the traction components:
\begin{subequations}
\begin{equation}
    T_1 Y_{l m} \equiv r^2 \delta \pi_r^{\hphantom{r} r} 
        = \frac{2 \mu}{3} [r^2 (K - H_2) - l (l + 1) V 
        - 2 r W' + (4 - r \lambda') W] Y_{l m}
\end{equation}
and
\begin{equation}
    T_2 \partial_\theta Y_{l m} \equiv r^3 \delta \pi_r^{\hphantom{r} \theta} 
        = - \mu (r V' - 2 V + e^\lambda W) \partial_\theta Y_{l m}.
\end{equation}
\end{subequations}

Using this information in the perturbed Einstein field equations 
(\ref{eq:PerturbedEinstein}) provides the elastic perturbation equations:
\begin{subequations}\label{eqs:ElasticPerturbations}
\begin{gather}
\begin{split}
    H_0'' + \left( \frac{2}{r} + \frac{\nu' - \lambda'}{2} \right) H_0' 
        + \left\{ \frac{2}{r^2} - [2 + l (l + 1)] \frac{e^\lambda}{r^2}
        + \frac{3 \nu' + \lambda'}{r} - \nu'^2 \right\} H_0 
        = - 8 \uppi \bigg[ &e^\lambda ( 3 \delta p + \delta \varepsilon )
        + 2 \nu' (\mu V)' \\
        &+ 8 \left( \frac{1 - e^\lambda}{r^2} + \frac{2 \nu' + \lambda'}{2 r} 
        - \frac{1}{4} \nu'^2 \right) \mu V + \frac{2 \nu'}{r} T_2 \bigg],
\end{split}\\
    K' = H_0' + \nu' H_0 + \frac{16 \uppi}{r} (2 + r \nu') \mu V 
        - \frac{16 \uppi}{r} T_2, \\
    W' - \left( \frac{2}{r} - \frac{\lambda'}{2} \right) W 
        = \frac{1}{2} r (K - H_0) 
        - \left[ 16 \uppi r \mu + \frac{l (l + 1)}{2 r} \right] V 
        - \frac{3}{4 r \mu} T_1, \\
    V' - \frac{2}{r} V = - \frac{e^\lambda}{r} W - \frac{1}{r \mu} T_2, \\
    T_2' + \left( \frac{\nu' - \lambda'}{2} + \frac{1}{r} \right) T_2 
        = - e^\lambda r \delta p + \frac{\nu' + \lambda'}{16 \uppi} H_0 
        + \frac{e^\lambda}{r} [2 - l (l + 1)] \mu V 
        + \frac{e^\lambda}{2 r} T_1, \\
	16 \uppi e^\lambda r^2 \delta p = r^2 \nu' H_0' 
        + [l (l + 1) e^\lambda - 2 + r^2 \nu'^2] H_0 
        + [2 - l (l + 1)] e^\lambda K + 16 \uppi r^2 \nu'^2 \mu V 
        - 16 \uppi e^\lambda T_1 - 16 \uppi (2 + r \nu') T_2, \\
    \frac{3}{4 \mu} T_1 
        = \frac{r^2}{(\varepsilon + p)} \delta \varepsilon 
        + \frac{3}{2} r^2 K - \frac{3}{2} l (l + 1) V 
        + \left(3 - \frac{r \nu'}{2 c_\text{s}^2}\right) W, \\
    H_1 = 0
\end{gather}
and 
\begin{equation}
    H_2 = H_0 + 32 \uppi \mu V.
\end{equation}
\end{subequations}
Equations~(\ref{eqs:ElasticPerturbations}) form a coupled system of equations 
that describe the static, even-parity perturbations in the elastic crust. It is 
straightforward to show that, when the shear modulus vanishes, 
equations~(\ref{eqs:ElasticPerturbations}) reduce to the perturbation 
equations~(\ref{eqs:FluidPerturbations}) for the fluid.

\subsection{Boundary conditions}

To solve the perturbation equations, we need to consider the boundary conditions 
of the problem. At the centre of the star, the solution must be regular, 
\begin{equation}
    H_0(0) = 0.
\end{equation}
However, as the structure equations are singular at the origin, we must begin 
integrating at small $r$. Regularity provides the initial condition, 
\begin{equation}
    H_0(r) = a_0 r^l [1 + \mathcal{O}(r^2)],
\label{eq:Initial}
\end{equation}
where $a_0$ is a constant that depends on the amplitude of the perturbations.

There are two fluid-elastic interfaces in star B: the core-crust and crust-ocean 
boundaries. Because the star is unstrained in the background, the background 
quantities will be continuous across these boundaries. However, should the 
equation of state involve discontinuities at the phase transitions, these would 
need to be accounted for \citep[e.g., see][]{2020ApJ...895...28P}.

In order to determine how the perturbed variables behave at a boundary, one 
must inspect the first and second fundamental forms, which both must be 
continuous at the interfaces 
\citep{1990MNRAS.245...82F, 2011PhRvD..84j3006P, 2020PhRvD.101j3025G}. 
From this analysis, one finds that the functions $H_0$, $K$ and $W$ must be 
continuous, along with the traction, which has components, 
$(T_1 + r^2 \delta p)$ and $T_2$. We assume the shear modulus has a finite value 
throughout the crust and, therefore, is discontinuous at a fluid-elastic 
boundary, since a fluid has a vanishing shear modulus. This is an important, yet 
subtle, point as in the alternative case, where the shear modulus approaches 
zero at an interface, one finds that the tangential traction condition is 
trivially satisfied and there are insufficient boundary conditions to solve the 
problem.

Towards the surface of both stars A and B, there is a fluid ocean. In a fluid, 
all perturbed metric functions and their derivatives are continuous. Therefore, 
the interior solution will match smoothly to the exterior. In the vacuum, we 
find that (\ref{eq:H_0Fluid}) reduces to
\begin{equation}
    H_0'' + \left( \frac{2}{r} - \lambda' \right) H_0'
        - \left[ l (l + 1) \frac{e^\lambda}{r^2} + \lambda'^2 \right] H_0 = 0.
\label{eq:H_0ExteriorODE}
\end{equation}
This is the relativistic analogue of the linearised Laplace's equation for 
non-radial perturbations. The solution to (\ref{eq:H_0ExteriorODE}) may be 
expressed in terms of associated Legendre polynomials and for quadrupolar 
($l = 2$) perturbations is
\begin{equation}
    H_0(r) = c_1 \left( \frac{r}{M} \right)^2 \left( 1 - \frac{2 M}{r} \right) 
        \left[- \frac{M (M - r) (2 M^2 + 6 M r - 3 r^2)}{r^2 (2 M - r)^2} 
        + \frac{3}{2} \ln \left( \frac{r}{r - 2 M} \right) \right] 
        + 3 c_2 \left( \frac{r}{M} \right)^2 \left( 1 - \frac{2 M}{r} \right), 
\label{eq:H_0Exterior}
\end{equation}
where $M \equiv m(R)$ is the total gravitational mass of the background and 
$c_1$ and $c_2$ are constants. From (\ref{eq:H_0Exterior}) we observe two 
solutions to the relativistic Laplace's equation: a decreasing solution with 
$c_1$, that is associated with the gravitational potential of the star, and an 
increasing solution with $c_2$, that may be associated with the tidal potential 
\citep[see, e.g.,][]{2008ApJ...677.1216H}. For other applications, one can 
assume $c_2$ = 0 and $H_0$ only contains the star's potential. By comparing 
(\ref{eq:MultipoleRelativity}) and (\ref{eq:H_0Exterior}), we identify
\begin{equation}
    Q_{2 m} = \frac{M^3 c_1}{\uppi},
\label{eq:Quadrupole}
\end{equation}
in agreement with equation~(41) in \citet{2013PhRvD..88d4004J}. Thus, one can 
obtain the quadrupole moment by identifying the decreasing solution in $H_0$.

In the fluid regions, one must solve for $(H_0', H_0)$ using 
equations~(\ref{eqs:FluidPerturbations}). In the elastic crust of star B, the 
perturbations are governed by equations~(\ref{eqs:ElasticPerturbations}) with 
variables $(H_0', H_0, K, W, V, T_2)$. Equation~(\ref{eq:Initial}) provides the 
initial boundary condition in the fluid core. The crust of star B presents a 
boundary-value problem with six boundary conditions: continuity of $H_0$ and $K$ 
at the core-crust interface and continuity of the traction at both interfaces. 
[The calculation of this boundary-value problem is described in detail in 
Appendix~B of \citet{2020PhRvD.101j3025G}.] At the surface, the solution must 
match (\ref{eq:H_0Exterior}).

\section{The source of the deformation}
\label{sec:Source}

For the pressure-density relation, we first use the analytic BSk24 equation of 
state \citep{2018MNRAS.481.2994P} for the high-density regions 
($\varepsilon > \SI{e6}{\gram\per\centi\metre\cubed}$) of the star and the 
\citet{2001A&A...380..151D} table for the low-density regions 
($\varepsilon \leq \SI{e6}{\gram\per\centi\metre\cubed}$). In the crust, we need 
to prescribe the shear-modulus profile. We use the \citet{1990PhRvA..42.4867O} 
result,
\begin{equation}
    \mu = 0.1994 \left( \frac{4 \uppi}{3} \right)^{1/3} 
        \left( \frac{1 - X_\text{n}}{A} n_\text{b} \right)^{4/3} (e Z)^2,
\end{equation}
where $X_\text{n}$ is the free neutron fraction, $A$ and $Z$ are the atomic and 
proton numbers, respectively, $n_\text{b}$ is the baryon number density and 
$e$ is the fundamental electric charge. We neglect any phase transitions in the 
crust (as they would significantly complicate the calculation). For the 
nuclear-matter parameters, we use the BSk24 results for the inner crust and the 
HFB-24 model \citep*{2013PhRvC..88b4308G} along with the experimental data from 
the 2016 Atomic Mass Evaluation \citep{2017ChPhC..41c0003W} for the outer crust 
\citep[see Table~4 in][]{2018MNRAS.481.2994P}. The location of the core-crust 
transition is given by the BSk24 results and the crust-ocean transition is 
taken to be the lowest density in the outer-crust model. The complexity of this 
prescription is due to our attempt to use a consistent model for the 
neutron star physics.

We consider three sources for the perturbations: (i) a deforming potential that 
is a solution to the relativistic Laplace's equation, (ii) a thermal pressure 
perturbation and (iii) a thermal pressure perturbation that only acts outside 
the core. Sources (i) and (ii) are the relativistic equivalents of two forces we 
considered in our Newtonian calculation \citep{2021MNRAS.500.5570G}. In reality, 
the fiducial force will depend on the evolutionary history of the star. Since 
this is a complex problem (beyond the scope of this work), the examples we 
consider are indicative in nature and should serve as illustrations of how one 
can calculate mountains using this scheme. The forces we use should not be 
interpreted as having an explicit link with the neutron star physics. To 
calculate the mountains for each example, we follow the 
\citet{2021MNRAS.500.5570G} scheme outlined in Section~\ref{sec:Scheme}. The 
perturbations are normalised by ensuring star B reaches breaking strain at a 
point in its crust. In practice, this means we take the point in the crust where 
the strain is greatest and set that to breaking strain. To obtain the 
perturbation amplitude, we use the von Mises criterion. In the formalism of 
\citet{2019CQGra..36j5004A}, the von Mises strain for a star with an unstrained 
background is
\begin{equation}
    \Theta = \sqrt{\frac{3}{2} \Delta s_{a b} \Delta s^{a b}}.
\end{equation}
The von Mises criterion states that a point in the crust fractures when the 
von Mises strain reaches the threshold, $\Theta \geq \Theta^\text{break}$, at 
some point. For $(l, m) = (2, 2)$ perturbations, we have the explicit expression
\begin{equation}
\begin{split}
    \Theta^2 = \frac{45}{512 \uppi} \Bigg\{ \sin^2 \theta \Bigg[ 
        &3 \sin^2 \theta \cos^2 2 \phi \left( \frac{T_1}{r^2 \mu} \right)^2 
        + 4 e^{-\lambda} (3 + \cos 2 \theta - 2 \sin^2 \theta \cos 4 \phi) 
        \left( \frac{T_2}{r^2 \mu} \right)^2 \Bigg] \\
        &+ (35 + 28 \cos 2 \theta + \cos 4 \theta + 8 \sin^4 \theta \cos 4 \phi) 
        \left( \frac{V}{r^2} \right)^2 \Bigg\}.
\end{split}
\label{eq:vonMisesStrain}
\end{equation}
Following the results of molecular-dynamics simulations by 
\citet{2009PhRvL.102s1102H}, we take the breaking strain to be 
$\Theta^\text{break} = 0.1$. Other estimates for the magnitude of the breaking 
strain include \citet{2018MNRAS.480.5511B} who obtained the smaller value of 
$\Theta^\text{break} = 0.04$. However, as was the case for previous 
maximum-mountain calculations, our results are linear in the breaking strain. 
Thus, a smaller breaking strain would result in less pronounced mountains.

\begin{figure}
    \centering
	\includegraphics[width=0.65\columnwidth]{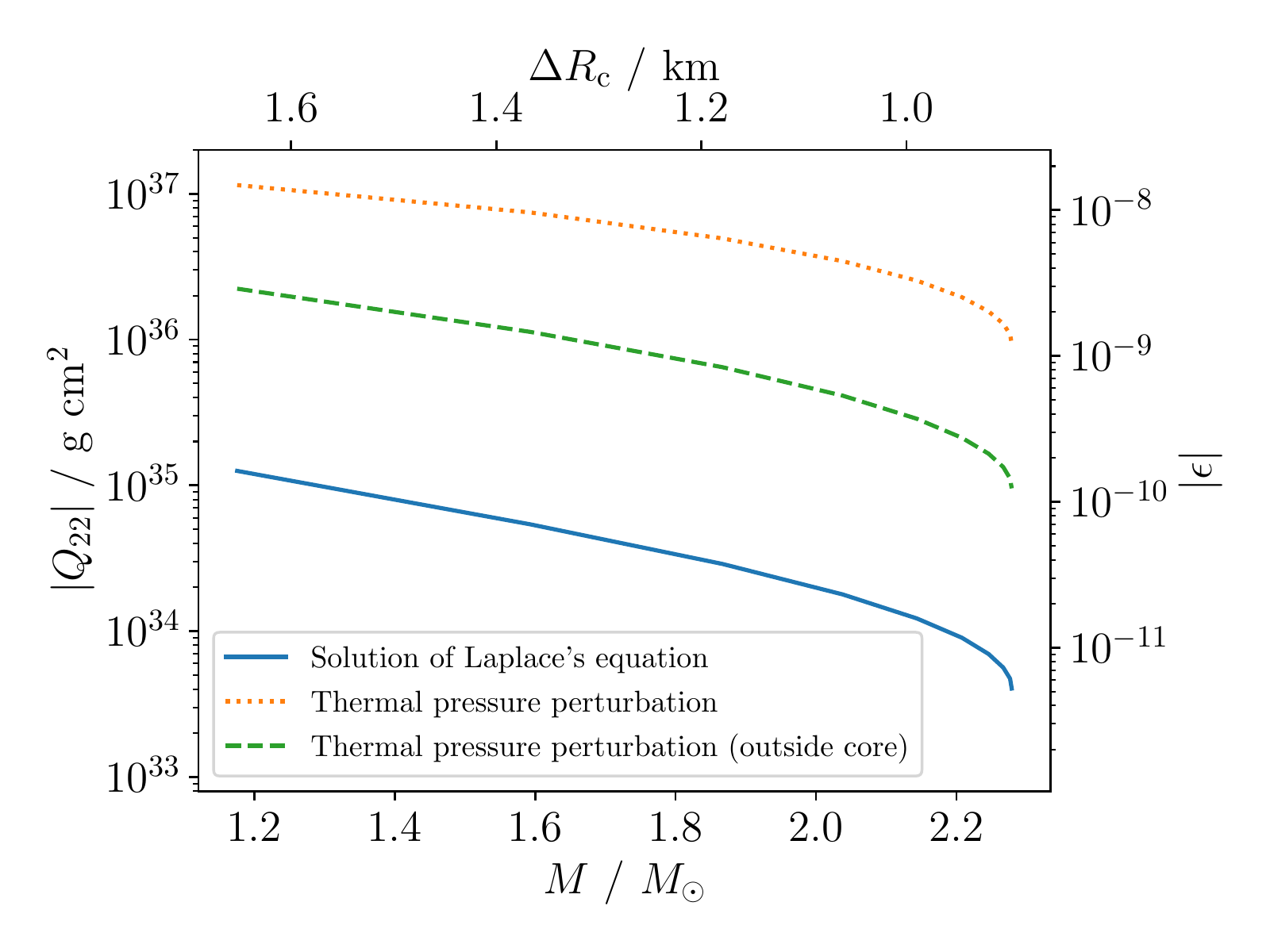}
    \caption{\label{fig:forces}The maximum quadrupole and ellipticity due to 
             the different forces as functions of stellar mass and crustal 
             thickness, $\Delta R_\text{c}$. We show results for the force 
             corresponding to the solution of Laplace's equation (solid blue 
             line), the thermal pressure perturbation (dotted orange line) and 
             the thermal pressure perturbation that acts outside the core 
             (dashed green line).}
\end{figure}

For our results, we will report, alongside the quadrupole moment, the fiducial 
ellipticity,
\begin{equation}
    \epsilon = \sqrt{\frac{8 \uppi}{15}} \frac{Q_{2 2}}{I_{z z}},
\end{equation}
where $I_{z z}$ is the principal stellar moment of inertia, taken to be 
$I_{z z} = \SI{e45}{\gram\centi\metre\squared}$. It is worth emphasising that 
the fiducial ellipticity does not perfectly correspond to the star's actual 
deformation. However, the principal moment of inertia is not expected to differ 
from the fiducial value by more than a factor of a few. We include the 
fiducial ellipticity to streamline comparison with observational results. Our 
results for the different forces are summarised in Fig.~\ref{fig:forces}.

\subsection{A solution to the relativistic Laplace's equation}

The relativistic Laplace's equation is given by the vacuum Einstein equations 
(\ref{eq:H_0ExteriorODE}). A common example of a solution to these equations is 
the tidal potential. In this situation, the perturbed metric function, $H_0$, 
has two contributions: the gravitational potential of the deformed star and the 
tidal potential of the companion. From (\ref{eq:H_0Exterior}), we can calculate 
the constants, $c_1$ and $c_2$, using $H_0$ and $H_0'$ in the exterior:
\begin{subequations}
\begin{equation}
    c_1 = \frac{r (r - 2 M)}{8 M^3} [r (2 M - r) H_0'(r) + 2 (r - M) H_0(r)]
\label{eq:PotentialAmplitude}
\end{equation}
and
\begin{equation}
\begin{split}
    c_2 = \frac{1}{48 M^3 r (2 M - r)} \Bigg\{
        - 2 M [r (4 M^4 + 6 M^3 r - 22 M^2 r^2 
        &+ 15 M r^3 - 3 r^4) H_0'(r) 
        + 2 (2 M^4 - 2 M^3 r + 13 M^2 r^2 
        - 12 M r^3 + 3 r^4) H_0(r)] \\
        &+ 3 r^2 (r - 2 M)^2 [r (2 M - r) H_0'(r) 
        + 2 (r - M) H_0(r)] \ln\left( \frac{r}{r - 2 M} \right) \Bigg\}.
\end{split}
\label{eq:ForceAmplitude}
\end{equation}
\end{subequations}
At the surface, $H_0$ and its derivative are continuous, so we can use their 
values at this point to calculate the force amplitude (\ref{eq:ForceAmplitude}) 
and the quadrupole (\ref{eq:Quadrupole}).

Therefore, we compute stars A and B using the fluid 
(\ref{eqs:FluidPerturbations}) and elastic perturbations equations 
(\ref{eqs:ElasticPerturbations}) for $(l, m) = (2, 2)$ perturbations. At the 
surface, we obtain the amplitude of the force from (\ref{eq:ForceAmplitude}) and 
increase its amplitude until a point in the crust breaks according to 
(\ref{eq:vonMisesStrain}). Then, once the two stars are normalised to the same 
force, we calculate the quadrupole moment with (\ref{eq:Quadrupole}) and 
(\ref{eq:PotentialAmplitude}).

Fig.~\ref{fig:forces} shows the maximum deformations with such a force for 
varying stellar mass. In the equivalent Newtonian case, for a star with 
$M = \SI{1.4}{\solarMass}$, $R = \SI{10}{\kilo\metre}$, we found 
$Q_{2 2} = \SI{1.7e37}{\gram\centi\metre\squared}$, 
$\epsilon = \num{2.2e-8}$ \citep{2021MNRAS.500.5570G}. We see that the 
corresponding maximum deformation for an $M = \SI{1.4}{\solarMass}$ star is 
two orders of magnitude lower. This suppression has two contributions. In going 
from Newtonian to relativistic gravity, the maximum size of the quadrupole that 
a crust can support decreases. This suppression was observed by 
\citet{2013PhRvD..88d4004J} in their relativistic calculation and has also been 
seen in tidal- 
\citep{2008ApJ...677.1216H, 2010PhRvD..81l3016H, 2009PhRvD..80h4035D, 
2009PhRvD..80h4018B} 
and magnetic-deformation calculations 
\citep{2004ApJ...600..296I, 2010MNRAS.406.2540C, 2012PhRvD..86d4012Y, 
2012MNRAS.427.3406F}. 
This behaviour has been attributed to the stiffness of the external, vacuum 
spacetime, that suppresses the quadrupole in the matching at the stellar surface 
\citep{2013PhRvD..88d4004J}.

\begin{figure}
    \centering
	\includegraphics[width=0.65\columnwidth]{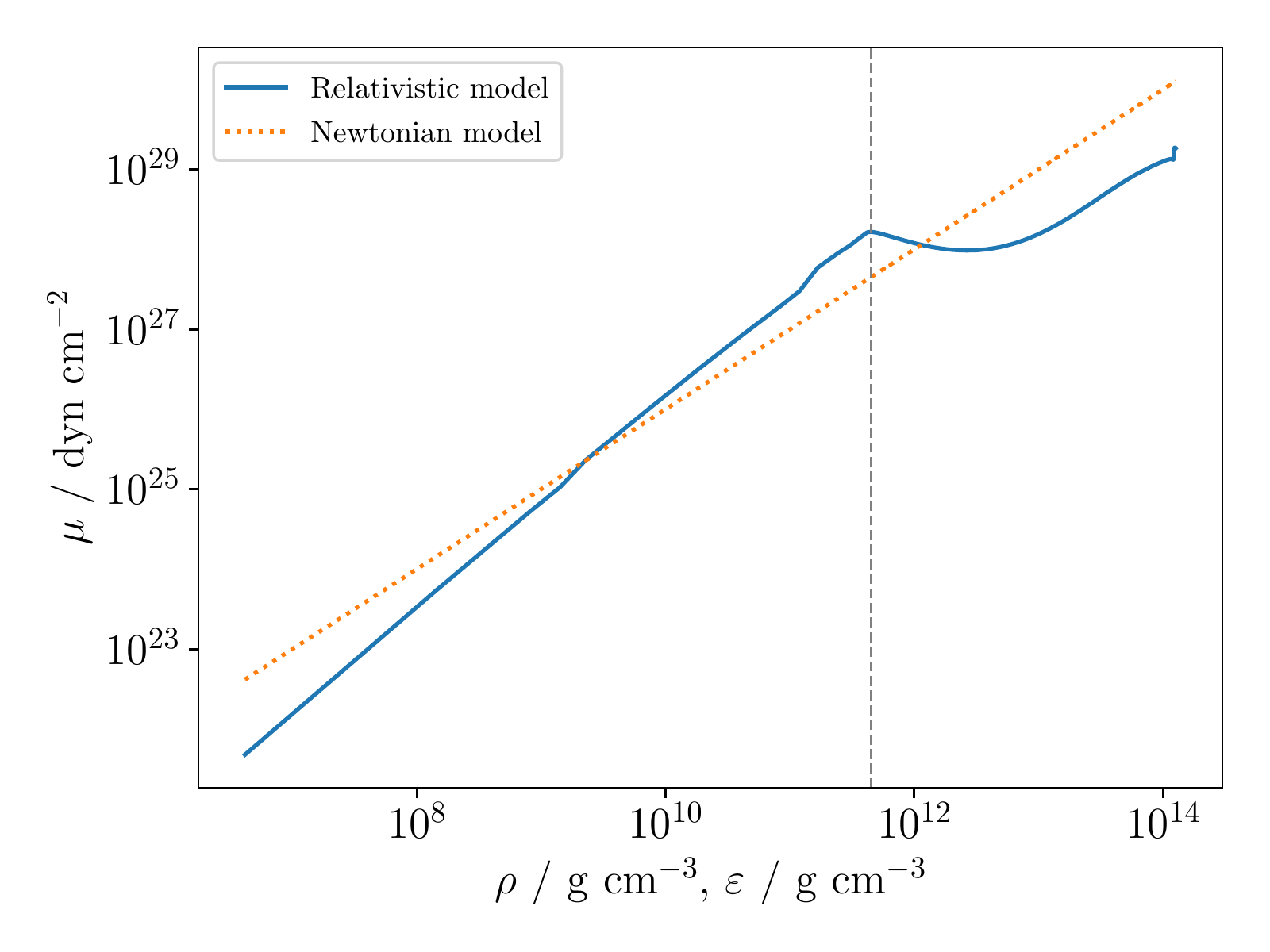}
    \caption{\label{fig:models}The shear modulus as a function of energy 
             density, $\varepsilon$, for the model of the crust in this 
             relativistic calculation (solid blue line) and as a function of 
             mass density, $\rho$, for the linear model used in the Newtonian 
             calculation (dotted orange line). We show the density at which the 
             crust transitions from the inner crust to the outer crust (known as 
             neutron drip; vertical dashed grey line).}
\end{figure}

The second effect comes from the equation of state. Focusing on the role of the 
matter model in our calculations, we found that the point where the strain was 
the largest for all the forces we considered was the top of the crust. This is 
where the crust yields which is consistent with previous calculations of neutron 
star crusts \citep{2020PhRvD.101j3025G, 2021MNRAS.500.5570G}. In 
Fig.~\ref{fig:models}, we compare the shear-modulus profile used in this work 
with the linear model used in our Newtonian calculation 
\citep{2021MNRAS.500.5570G}. Although the linear model appears to be a 
reasonable approximation to the more realistic model, there are key areas where 
the two differ. Of particular importance to the maximum-mountain calculation, 
the realistic shear-modulus profile is significantly weaker in the lower-density 
outer crust (approximately an order of magnitude smaller at the top of the 
crust). This plays a pivotal role in determining the size of the mountains that 
the crust can support as the breaking strain scales with the shear modulus.

Our results are at least two to three orders of magnitude smaller than the 
mountains obtained by \citet{2013PhRvD..88d4004J}. As was noted by 
\citet{2021MNRAS.500.5570G}, the fact that our scheme produces smaller mountains 
than calculated in previous work is not particularly surprising. Indeed, the 
very nature of the \citet{2000MNRAS.319..902U} approach [that 
\citet{2013PhRvD..88d4004J} follow] is to ensure that the \textit{entire} crust 
is at breaking strain, whereas in our scheme (in order to correctly satisfy the 
boundary conditions of the problem) breaking strain is reached at a point. 
Clearly, the size of the mountains depend on the force, so it is natural to 
explore different choices to see if we can obtain larger deformations. We will 
go on to consider a couple of other examples.

\subsection{A thermal pressure perturbation}
\label{subsec:Thermal}

The next source for the perturbations enters the problem in the same way as a 
thermal pressure. For this case, we assume that the pressure has a thermal 
component of the form \citep{2021MNRAS.500.5570G},
\begin{equation}
    \delta p_\text{th} = \frac{k_\text{B} \varepsilon}{m_\text{u}} \delta T,
    \label{eq:ThermalPressure}
\end{equation}
where $k_\text{B}$ is the Boltzmann constant, $m_\text{u}$ is the atomic mass 
unit and $\delta T$ is the perturbation to the temperature. Since the 
temperature perturbation must be regular at the centre of the star, we will 
assume its radial dependence is given by
\begin{equation}
    \delta T(r) = \left( \frac{r}{R} \right)^2 \delta T(R).
\label{eq:Temperature}
\end{equation}
To incorporate this force into the perturbation 
equations~(\ref{eqs:FluidPerturbations}) and (\ref{eqs:ElasticPerturbations}), 
they must be adjusted by $\delta p \rightarrow \delta p + \delta p_\text{th}$. 
In this case, $H_0$ corresponds solely to the gravitational potential of the 
star.

For this example, the amplitude $a_0$ in (\ref{eq:Initial}) is connected to the 
magnitude of the temperature perturbation. It is constrained by ensuring the 
function $H_0$ matches the exterior solution (\ref{eq:H_0Exterior}) with 
$c_2 = 0$. This boundary condition can be expressed as
\begin{equation}
    - \frac{1}{2} (R - 2 M) H_0'(R) = \left\{ - 1 + \frac{M}{R} 
        - \frac{8 M^5}{2 M R (M - R) (2 M^2 + 6 M R - 3 R^2) + 3 R^3 (R - 2 M)^2 
        \ln[(R - 2 M) / R]} \right\} H_0(R),
\label{eq:Surface}
\end{equation}
which is the relativistic analogue (for $l = 2$ perturbations) of the Newtonian 
boundary condition at the surface in (\ref{eq:BoundaryConditions}). The 
temperature perturbation, $\delta T(R)$, is then increased until a point in the 
crust breaks.%
\footnote{For a canonical $M = \SI{1.4}{\solarMass}$ star, we found the 
crust breaks when $\delta T(R) = \SI{9.4e4}{\kelvin}$. The temperature reported 
here is not a physical temperature perturbation the star is subjected to, noting 
that the background is at zero temperature. It is simply a source term for the 
pressure perturbation~(\ref{eq:ThermalPressure}).}

The maximum mountains built using this force are shown in Fig.~\ref{fig:forces}. 
Compared to the solution to the relativistic Laplace's equation, the thermal 
pressure perturbation produces mountains approximately two orders of magnitude 
larger. As we saw in the Newtonian calculation \citep{2021MNRAS.500.5570G}, this 
illustrates that the size of the mountains that can be built are highly 
dependent on their formation history.

For comparison, the corresponding force in our Newtonian calculation gave 
$Q_{2 2} = \SI{4.0e38}{\gram\centi\metre\squared}$, $\epsilon = \num{5.2e-7}$. 
Hence, the suppression for the thermal pressure is weaker than in the previous 
example.

We also considered the situation where the top of the crust was moved to neutron 
drip in order to assess the impact of removing the weaker regions of the crust. 
In this case, for an $M = \SI{1.4}{\solarMass}$ star, we obtained 
$Q_{2 2} = \SI{5.9e38}{\gram\centi\metre\squared}$, $\epsilon = \num{7.7e-7}$, 
which is even larger than what was obtained in the Newtonian calculation. This 
is not particularly surprising since we only focus on the inner crust, which is 
orders of magnitude stronger in the shear modulus than most of the outer crust 
(see Fig.~\ref{fig:models}). We also found that the crust no longer yielded at 
the top. This illustrates the role the shear modulus plays in supporting the 
mountains.

\subsection{A thermal pressure perturbation outside the core}

We also consider the case where the thermal pressure perturbation does not reach 
into the core. One could imagine a scenario where the surface of the neutron 
star is heated and this heating does not penetrate to the core. We will assume 
that the thermal pressure has a finite value at the base of the crust and exists 
in the crust and ocean. We use the same form for the temperature perturbation 
(\ref{eq:Temperature}).

Due to the specificity of this force, one must be careful in setting up the 
calculation. Since the core is unperturbed, we will assume 
$H_0(r_\text{base}) = W(r_\text{base}) = 0$, where $r_\text{base}$ denotes the 
position of the base of the crust. Because the force suddenly appears at the 
base, we assume that $H_0'(r_\text{base})$ is non-zero. For the fluid star 
(star A), this is sufficient to calculate the mountains. The precise value of 
$H_0'$ at the base is determined by ensuring the surface boundary condition 
(\ref{eq:Surface}) is satisfied.

For the star with a crust (star B), we need to pay attention to the traction 
conditions. Since $H_0'$ needed to have a finite value at the base of the crust 
in the fluid star, we effectively violated the radial traction condition 
[cf. (A17) in \citet{2020PhRvD.101j3025G}]. However, we can still use the 
tangential traction condition to demand $T_2(r_\text{base}) = 0$. This leaves 
two free values, $K(r_\text{base})$ and $V(r_\text{base})$. At the top of the 
crust we still impose both traction conditions, which constrains 
$K(r_\text{base})$ and $V(r_\text{base})$ and, thus, the problem is well posed. 
As was the case for the fluid star, $H_0'(r_\text{base})$ is constrained via 
(\ref{eq:Surface}).

We plot the maximum mountains for this case in Fig.~\ref{fig:forces}. As 
compared to the thermal pressure that acts throughout the star, the mountains 
produced in this example are approximately an order of magnitude smaller.

In summary, we have seen with all our examples in Fig.~\ref{fig:forces} that the 
maximum deformations the crust can sustain are small. Nevertheless, it is 
interesting to note that our results are, in principle, large enough to be 
consistent with the minimum deformation argument of \citet{2018ApJ...863L..40W} 
and the quadrupoles that can describe the accreting millisecond pulsar 
population \citep{2019MNRAS.488...99G}.

\section{Dependence on the equation of state}
\label{sec:Dependence}

As the calculation is done in full general relativity, we have the opportunity 
to assess the impact of the equation of state. To do this, we explore a subset 
of the chiral effective-field-theory models combined with a speed-of-sound 
parametrisation \citep[see][]{2018ApJ...860..149T}. [These models were recently 
used by \citet{2020NatAs...4..625C} to obtain constraints on neutron star 
radii.] Chiral effective field theory is a systematic framework for low-energy 
hadronic interactions. For low densities, the theory describes matter using 
nucleons and pions, where the interactions are expanded in powers of momenta and 
all the relevant operators in strong interactions are included 
\citep{1990PhLB..251..288W, 1991NuPhB.363....3W, 1994PhRvC..49.2932V, 
2009RvMP...81.1773E, 2011PhR...503....1M}. 
One then uses quantum Monte Carlo methods to solve the many-body Schr{\"o}dinger 
equation to obtain an equation of state 
\citep{2010PhRvC..82a4314H, 2013PhRvC..88b5802K, 2015RvMP...87.1067C, 
2018ApJ...860..149T}. 
This approach is expected to describe matter well up to between one to two times 
nuclear saturation density. \citet{2018ApJ...860..149T} extended the equations 
of state to higher densities using a speed-of-sound parametrisation to ensure 
that causality was not violated.

\begin{figure}
    \includegraphics[width=0.49\columnwidth]{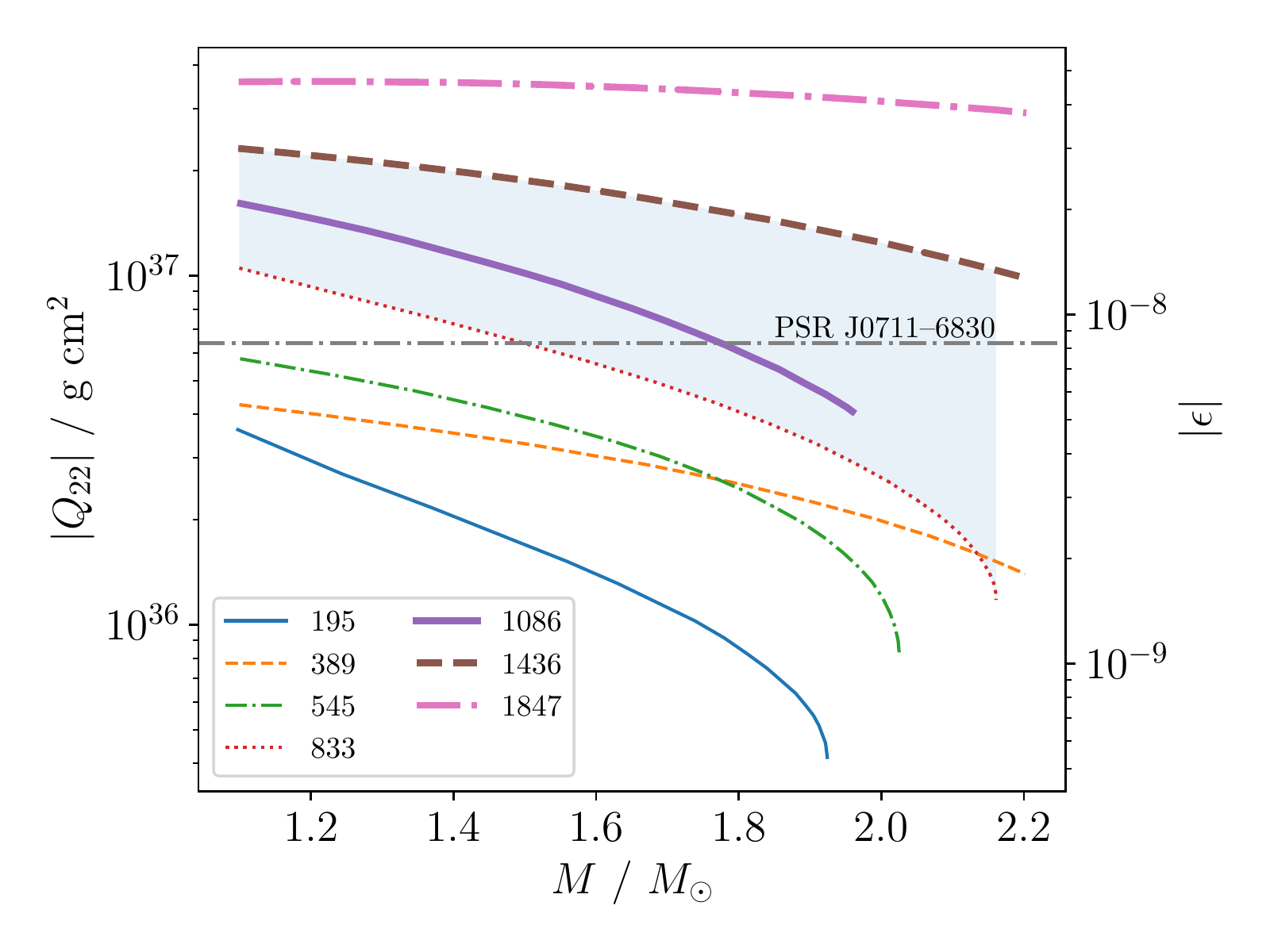}
	\includegraphics[width=0.49\columnwidth]{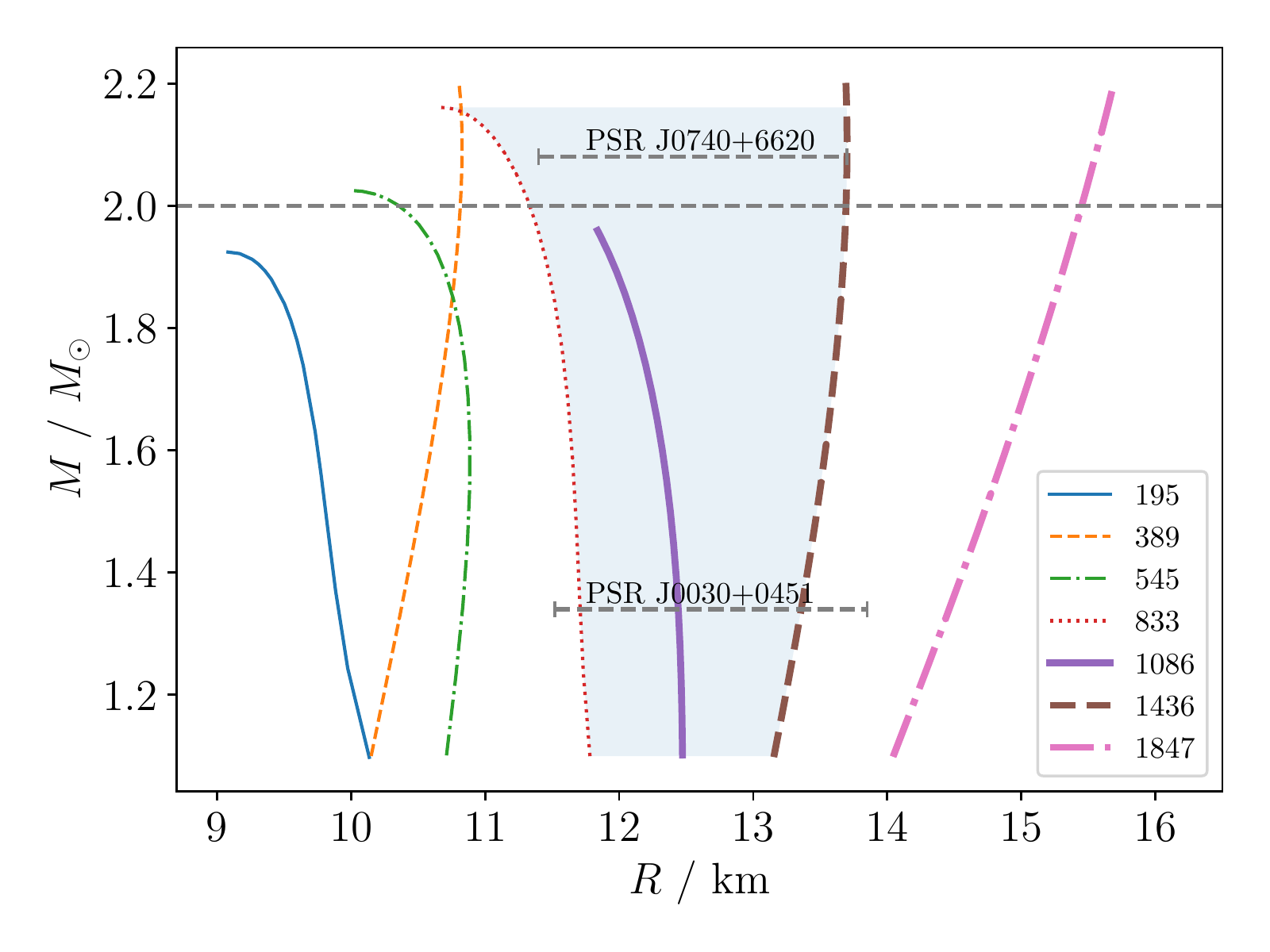}
    \caption{\label{fig:chiral}The maximum quadrupole and ellipticity due to 
             thermal pressure perturbations as functions of stellar mass for 
             different chiral effective-field-theory equation-of-state models 
             (left panel) and the corresponding mass-radius diagram traced out 
             by the background stellar models (right panel). To compare with 
             observational constraints, we indicate the $M = \SI{2}{\solarMass}$ 
             constraint (middle dashed grey line) and the range of radii 
             $11.52 \leq R \ / \ \si{\kilo\metre} \leq 13.85$ 
             (bottom dashed grey line) and 
             $11.4 \leq R \ / \ \si{\kilo\metre} \leq 13.7$ 
             (top dashed grey line) measured for PSR J0030+0451 and PSR 
             J0740+6620, respectively, in the right panel. We shade the region 
             between the two models 833 and 1436 that (roughly) satisfy the 
             constraints to give an indication of the range of possible maximum 
             deformations for this force. On the left panel, we indicate the 
             upper limit on the ellipticity of PSR J0711--6830, 
             $\epsilon < \num{8.3e-9}$ (dash-dotted grey line). The equations of 
             state are indexed according to their radii for an 
             $M = \SI{1.4}{\solarMass}$ star. All stellar models considered are 
             stable to radial perturbations.}
\end{figure}

We consider a selection of models for the pressure-density relation 
[supplemented by the \citet{2001A&A...380..151D} table for the low-density 
regions ($\varepsilon \leq \SI{e6}{\gram\per\centi\metre\cubed}$)] and subject 
the stars to thermal pressure perturbations (as described in 
Sec.~\ref{subsec:Thermal}). We choose this mechanism since it produced the 
largest mountains from the examples we considered. The results are shown in 
Fig.~\ref{fig:chiral}.

It should be noted that there are observational and theoretical constraints on 
the mass and radius of neutron stars. From observations, it is apparent that 
the true nuclear-matter equation of state must be able to support 
\SI{2}{\solarMass} neutron stars \citep{2013Sci...340..448A}. 
Additionally, there are recent constraints on the radius from NICER: 
PSR J0030+0451, with mass $M = \SI{1.34}{\solarMass}$, was measured to have 
$11.52 \lesssim R \ / \ \si{\kilo\metre} \lesssim 13.85$ 
\citep{2019ApJ...887L..21R} and PSR J0740+6620, with mass 
$M = \SI{2.08}{\solarMass}$, was measured to have 
$11.41 \lesssim R \ / \ \si{\kilo\metre} \lesssim 13.69$ 
\citep{2021arXiv210506980R}. 
There have also been studies combining observations and theory to constrain the 
radius of a canonical $M = \SI{1.4}{\solarMass}$ neutron star. 
\citet{2020ApJ...893L..21R} used the NICER observation of PSR J0030+0451 along 
with GW170817 to obtain a constraint of 
$11.75 \lesssim R \ / \ \si{\kilo\metre} \lesssim 13.5$. However, there is some 
degree of statistical uncertainty associated with this range and such 
constraints will need to be updated with future detections. Indeed, the study of 
\citet{2020NatAs...4..625C}, that combined nuclear theory with observations of 
GW170817, found the contrasting (and more stringent) range 
$10.4 \lesssim R \ / \ \si{\kilo\metre} \lesssim 11.9$ for an 
$M = \SI{1.4}{\solarMass}$ neutron star. We add the mass limit of 
$M = \SI{2}{\solarMass}$ and the radius ranges from the NICER measurements to 
the right panel of Fig.~\ref{fig:chiral}. The majority of the equations of state 
that we consider support stars with $M = \SI{2}{\solarMass}$. To give some 
indication of the accepted range of equations of state in the mass-radius 
diagram, we shade the region between models 833 and 1436, which (roughly) 
satisfy the observational constraints. In order to give context for some of the 
most constrained upper limits from gravitational-wave data, we show the 
deformation constraint for PSR J0711--6830 of $\epsilon < \num{8.3e-9}$ in the 
left panel \citep{2020ApJ...902L..21A}.

In the left panel of Fig.~\ref{fig:chiral}, we show the maximum deformations 
due to thermal pressure perturbations for the different equations of state. 
It is perhaps surprising to observe that there is a range of approximately two 
orders of magnitude across all the models. Thus, these mountain calculations 
are quite sensitive to the equation of state. However, as indicated by the 
shaded region of Fig.~\ref{fig:chiral}, the range of maximum deformations is 
narrower when we consider recent observational constraints on the equation of 
state.

One can also see that there is a relationship between the radius, $R$, of a 
star with a given mass, $M$, and the maximum deformation it supports. This 
should not be surprising, since equations of state that produce stars with 
larger radii will also have thicker crusts that can support larger deformations.

\section{Conclusions}
\label{sec:Conclusions}

There is hope on the horizon that we will soon detect gravitational waves from 
rotating neutron stars for the very first time. Indeed, there are continued 
efforts to improve the data-analysis techniques 
\citep[see, e.g.,][]{2021PhRvD.103f4027B, 2020PhRvL.125q1101D, 
2021PhRvD.103f3019D, 2021ApJ...909...79S, 2021ApJ...906L..14Z}, 
along with plans in development for third-generation gravitational-wave 
detectors to be constructed in the (hopefully) not-too-distant future 
\citep{2020JCAP...03..050M}. It is, therefore, meaningful to ask what may be the 
largest mountain neutron star crusts can support. Furthermore, such an answer 
could provide some helpful context in understanding the spin distribution of 
observed pulsars.

In our previous paper \citep{2021MNRAS.500.5570G}, we surveyed previous 
maximum-mountain calculations and found that there were issues relating to 
boundary conditions that must be satisfied for realistic neutron stars. In 
particular, the usual approach to calculating maximum neutron star mountains of 
\citet{2000MNRAS.319..902U} assumes a strain field that violates continuity of 
the traction. We introduced a new scheme for calculating mountains that gives 
one full control of the boundary conditions at the cost of requiring a knowledge 
of the deforming force that sources the mountains. However, it is unclear what 
this force should be. In reality, the force is related to the formation history 
of the star, that may involve complex mechanisms like quakes and accretion from 
a companion. Therefore, in order to get a handle on the force, evolutionary 
calculations that consider the history of neutron stars will be necessary. It 
should be noted that such calculations would need to be ambitious in order to 
take into account the physics that may be important in the evolution of the 
crust, such as cooling, freezing, spin down, magnetic fields and cracking (to 
name but a few mechanisms).

In this study, we generalised the scheme of \citet{2021MNRAS.500.5570G} to 
relativity. We considered three examples for the deforming force and found the 
most promising results for thermal pressure perturbations. In constructing 
relativistic stellar models with a realistic equation of state, we noted that 
the size of the mountains was suppressed (in some cases, quite significantly). 
This is generally due to two factors: (i) using relativistic gravity and (ii) 
the shear modulus of the crust is weaker at points than the simple model used 
in \citet{2021MNRAS.500.5570G}. For most of the examples we examined, the crust 
yielded first at the top, where the shear modulus is the weakest. These results 
also point towards the necessity of evolutionary calculations in making progress 
on mountain calculations. This is evident from the role the deforming force 
plays in how large the mountains can be.

We have also demonstrated how the mountains are sensitive to the 
pressure-density relationship for nuclear matter, suggesting a range of 
uncertainty about an order of magnitude for a variety of equations of state 
that satisfy current observational constraints on the mass and radius. For this 
analysis we considered a subset of equations of state from chiral effective 
field theory, that all obey causality, and subjected the stars to thermal 
pressure perturbations.

We made an effort to build a consistent neutron star model, based on the BSk24 
equation of state \citep{2018MNRAS.481.2994P}. However, our treatment of the 
crust is still somewhat simplistic: we assume that the crust behaves like 
an elastic solid up to some breaking strain, at which it yields and all the 
strain is subsequently released. As discussed by \citet{2021MNRAS.500.5570G}, 
a possible solution to constructing larger mountains than the ones we have been 
able to obtain perhaps lies in plasticity. Plastic solids behave elastically up 
to some strain and, beyond that strain, deform in such a way that they retain 
some of the strain. It is perhaps likely that neutron star crusts exhibits some 
of this behaviour. Suppose a neutron star crust is modelled as an ideal plastic. 
It is then deformed up to its elastic yield limit at a point in the crust and 
the strain saturates. Even if the force is increased, the strain stays the same. 
One could then continue to apply forces to the crust in order to build as large 
a mountain as possible. This may connect real neutron stars to the maximally 
strained configuration imposed in \citet{2000MNRAS.319..902U}. At present, this 
is speculation, even though there have been some interesting discussions of 
plasticity in neutron star crusts 
\citep{1970PhRvL..24.1191S, 2003ApJ...595..342J, 2010MNRAS.407L..54C}. This idea 
certainly seems worthy of future studies.

\section*{Acknowledgements}

NA is thankful for financial support from STFC via Grant No. ST/R00045X/1. The 
authors are especially grateful to I. Tews for supplying the chiral 
effective-field-theory equations of state.

\section*{Data availability}

Additional data underlying this article will be shared on reasonable request to 
the corresponding author.


\bibliographystyle{mnras}
\bibliography{bibliography}



\bsp	
\label{lastpage}
\end{document}